\documentclass[namedreferences]{solarphysics}

\usepackage[hyperref,optionalrh,solaromanenum]{spr-sola-addons}
\usepackage{graphicx}                    
\usepackage{amssymb}               
\usepackage{gensymb}
\usepackage{longtable}
\usepackage{color}                     
\usepackage{breakurl}                    
                      
\hypersetup{colorlinks, citecolor=blue, filecolor=black, linkcolor=black, urlcolor=black}

\usepackage{booktabs}
\usepackage{lscape}
\usepackage{siunitx}
\usepackage{pgfplotstable}

\makeatletter

\newcommand{\Rmnum}[1]{\expandafter\@slowromancap\romannumeral #1@}
\makeatother

\newcommand{\eg}{\emph{e.g.}}
\newcommand{\ie}{\emph{i.e.}}
\newcommand{\cf}{\emph{cf.}}

\newcommand{\corr}[1]{\textcolor{black}{#1}}
\newcommand{\corrtwo}[1]{\textcolor{black}{#1}}

\pgfplotstableset{col sep=comma, columns = {Flare_date,x,y,Flare_st,wave_time,Flare_class,
Active_Region,Num_consecutive_arcs,Central_arc_angle,Wave_present,median_vel,median_acc}}

\begin{document}

\begin{article}

\begin{opening}

\title{A Statistical Analysis of the Solar Phenomena Associated with Global EUV Waves}

\author[addressref=aff1,corref,email={david.long@ucl.ac.uk}]{\inits{D.M.}\fnm{D.M.}~\lnm{Long}\orcid{0000-0003-3137-0277}}
\author[addressref=aff2]{\inits{P.}\fnm{P.}~\lnm{Murphy}\orcid{0000-0002-8606-2645}}
\author[addressref=aff1]{\inits{G.}\fnm{G.}~\lnm{Graham}\orcid{0000-0002-3769-2587}}
\author[addressref=aff2]{\inits{E.P.}\fnm{E.P.}~\lnm{Carley}\orcid{0000-0002-6106-5292}}
\author[addressref=aff1]{\inits{D.}\fnm{D.}~\lnm{P\'{e}rez-Su\'{a}rez}\orcid{0000-0003-0784-6909}}

\runningauthor{D.M. Long \emph{et al.}}
\runningtitle{Statistical Analysis of the Solar Phenomena Associated with Global EUV Waves}

\address[id=aff1]{UCL-Mullard Space Science Laboratory, Holmbury St.~Mary, Dorking, Surrey, RH5~6NT, UK}
\address[id=aff2]{School of Physics, Trinity College Dublin, College Green, Dublin 2, Ireland}

\begin{abstract}
Solar eruptions are the most spectacular events in our solar system and are associated with many 
different signatures of energy release including solar flares, coronal mass ejections, global waves, 
radio emission and accelerated particles. Here, we apply the Coronal Pulse Identification and Tracking 
Algorithm (CorPITA) to the high cadence synoptic data provided by the \corr{\emph{Solar Dynamic Observatory}} (SDO) 
to identify and track global waves observed by SDO. 164 of the 362 solar flare events studied (45~\%) are 
found to have associated global waves with no waves found for the remaining 198 (55~\%). A clear linear 
relationship was found between the median initial velocity and the acceleration of the waves, with faster 
waves exhibiting a stronger deceleration (consistent with previous results). No clear relationship was 
found between global waves and type~\Rmnum{2} radio bursts, electrons or protons detected \emph{in-situ} 
near Earth. While no relationship was found between the wave properties and the associated flare size 
(with waves produced by flares from B to X-class), more than a quarter of the active regions studied were 
found to produce more than one wave event. These results suggest that the presence of a global wave in a solar 
eruption is most likely determined by the structure and connectivity of the erupting active region and the 
surrounding quiet solar corona rather than by the amount of free energy available within the active region.
\end{abstract}


\end{opening}

\section{Introduction}\label{s:intro} 

Global waves in the low solar corona (commonly called ``EIT waves'') were first observed using the Extreme 
ultraviolet Imaging Telescope \citep[EIT;][]{Dela:1995} onboard the \emph{Solar and Heliospheric Observatory} 
\citep[SOHO;][]{Domingo:1995}. Initially identified as fast-mode MHD waves, 
\citep[\eg,][]{Dere:1997,Moses:1997,Thompson:1998}, this interpretation was questioned following observations 
of stationary bright fronts at coronal hole boundaries and anomalously low measured kinematics 
\citep[\cf][]{Delannee:1999}. This led to the development of two distinct families of theories to describe this 
phenomenon; that they are alternatively waves (either linear or non-linear waves) or pseudo-waves (\ie, a 
brightening resulting from the restructuring of the coronal magnetic field during the eruption of a coronal mass 
ejection). Note that a more detailed overview of the different theories proposed to explain the ``EIT wave'' 
phenomenon may be found in the recent reviews by \citet{Liu:2014} and \citet{Warmuth:2015}. However, the advent 
of high-cadence observations with the launch of the \emph{Solar Terrestrial Relations Observatory} 
\citep[STEREO;][]{Kaiser:2008} and \emph{Solar Dynamics Observatory} \citep[SDO;][]{Pesnell:2012} 
spacecraft has begun to refine our understanding of this phenomenon. Recent work comparing the predictions made 
by each of these theories with observations suggests that they are best described as large-amplitude waves 
initially driven by the rapid lateral expansion of a CME in the low corona, before propagating freely 
\citep[\cf][]{Long:2017}. 

Although our understanding of the origin and physical properties of global waves has progressed since they were 
first observed, their relationship with other solar phenomena such as solar flares, CMEs, solar energetic particles 
(SEPs) and radio bursts continues to be a source of investigation. Global ``EIT waves'' have traditionally been 
studied using single event case-studies, making it difficult to draw general conclusions about the nature of their 
relationship with these phenomena. Recognising this issue, a catalogue of global ``EIT waves'' observed by SOHO/EIT 
was assembled by \citet{Thompson:2009}, with each wave event identified `by-eye' and classified using a quality 
rating system. This catalogue was subsequently used to investigate the link between global ``EIT waves'' and other 
solar phenomena including type~\Rmnum{2} radio bursts, solar flares and CMEs 
\citep[\eg,][]{Biesecker:2002,Warmuth:2011}. More recent work has extended this systematic approach to observations 
from the Extreme UltraViolet Imager \citep[EUVI;][]{Wuelser:2004} onboard STEREO \citep{Muhr:2014,Nitta:2014} and 
SDO/AIA \citep{Nitta:2013}. \corr{In each of these cases the global waves were identified using semi-automated techniques;  
\citet{Muhr:2014} defined the direction into which the wavefront propagated and used a perturbation profile technique 
to fit the leading edge of the wavefront while \citet{Nitta:2013,Nitta:2014} used 2-d intensity stack plots 
produced by a series of arc sectors to visually identify the leading edge of the wavefront. Each approach requires manual
input from the user, potentially making them susceptible to user bias.} In addition, the catalogues created using observations 
from SOHO/EIT and STEREO/EUVI may have been subject to the lower temporal resolution of both instruments, which 
could have led to a systematic under-estimation of the kinematics of the global waves \citep[\cf][]{Byrne:2013}.

Despite these issues, the catalogues developed by \citet{Thompson:2009} (in particular), \citet{Muhr:2014} and 
\citet{Nitta:2013} have been widely used to study the relationship between global waves and other solar phenomena 
such as type~\Rmnum{2} radio bursts, solar flares and CMEs. Although initially ambiguous, the relationship between 
global waves and CMEs is now well defined, with \citet{Biesecker:2002} showing that every wave has an associated CME, 
although not every CME has an associated wave. Type~\Rmnum{2} radio bursts have long been observed in the solar corona 
associated with solar eruptions \citep[\eg,][]{Payne:1947,Wild:1950} and it was generally accepted that both 
type~\Rmnum{2} bursts and Moreton-Ramsey waves \citep[first observed in the 1960's by][]{Moreton:1960a,Moreton:1960b} 
were signatures of the same driving process \citep{Uchida:1968}. However, the much lower measured speeds and other 
discrepancies between Moreton--Ramsey and global ``EIT waves'' complicated extending this assumption to ``EIT waves''. 
Instead, \citet{Klassen:2000} found that 90~\% of type~\Rmnum{2} bursts identified in 1997  were associated with 
``EIT waves''. In contrast, \citet{Biesecker:2002} used the wave event list compiled by \citet{Thompson:2009} to 
show that only 29~\% of the waves in the list had an associated type~\Rmnum{2} radio burst. This percentage was 
supported by an analysis of 60 global EUV waves observed by STEREO/EUVI and studied by \citet{Muhr:2014}, who found 
that 22~\% of the global waves studied had an associated type~\Rmnum{2} radio burst. However, a study of 138 global 
waves identified by \citet{Nitta:2013} using SDO/AIA found that 54~\% of the waves were associated with a 
type~\Rmnum{2} radio burst. The exact nature of the connection between global waves and type~\Rmnum{2} radio bursts 
therefore remains anomalous. 

The energy release during a solar eruption (either as a flare or through acceleration of a CME) can also result in 
solar energetic particles (SEPs) being accelerated into the heliosphere. SEPs are known to fall into two general 
categories, impulsive (typically associated with particle acceleration in a small area such as a solar flare) and 
gradual (associated with particle acceleration over a broad area, such as from a CME) as described by \citet{Reames:1993}. 
Release of gradual SEPs tends to occur close to the Sun, where the CME shocks as it propagates outwards into the 
heliosphere \citep[\eg,][]{Kahler:1994}. This process can also occur in the low corona, with the lateral expansion of 
the CME shock front accelerating SEPs \citep[\eg,][]{Rouillard:2012} as it propagates through the low corona. The result 
of this impulsive lateral expansion is best observed as a global EUV wave \citep[\cf.][]{Long:2017}. Despite this, there 
is no clear relationship between global EUV waves and SEP events, with most previous work tending to focus on individual 
case study events \citep[\eg,][]{Kozarev:2011,Prise:2014}. The measured SEP detection time is then typically used to 
infer the time and location that the particles were released on the Sun, which can be compared to the tracked evolution 
of the global wave in the low corona \citep[\eg,][]{Rouillard:2012,Prise:2014}. This approach was also used by 
\citet{Miteva:2014} to study 179 SEP events between 1997 and 2006, finding that protons detected in-situ were related 
to the global EUV waves, but that there was no correlation between the waves and electrons.

Here we describe the application of an automated algorithm to observations from SDO/AIA to identify and characterise 
global EUV waves and relate them to other solar phenomena such as flares, CMEs, SEP events and type~\Rmnum{2} 
radio bursts. The various data-sets and how the measurements were made are described in Section~\ref{s:data}, with the 
results of the statistical analysis described in Section~\ref{s:results}. The results are then discussed and some 
conclusions drawn in Section~\ref{s:disc}. Note that the complete table used for the analysis described here is also 
included in the Appendix.

\section{Observations and Data Analysis}\label{s:data}

The list of global EUV waves identified by \citet{Nitta:2013} was used as a starting point for this investigation to 
maximise the number of global wave events and associated phenomena that could be studied in detail. In each case, 
the location of the flare associated with each eruption was used as the source of the global wave, with the start time 
of the flare used as the reference point for analysing the EUV, radio and in-situ data. 

\subsection{Global Wave Characterisation and Analysis}\label{ss:waves}

\begin{figure*}[!t]
\centering{
	\includegraphics[width=0.99\textwidth,clip=,trim=0mm 0mm 14mm 0mm]{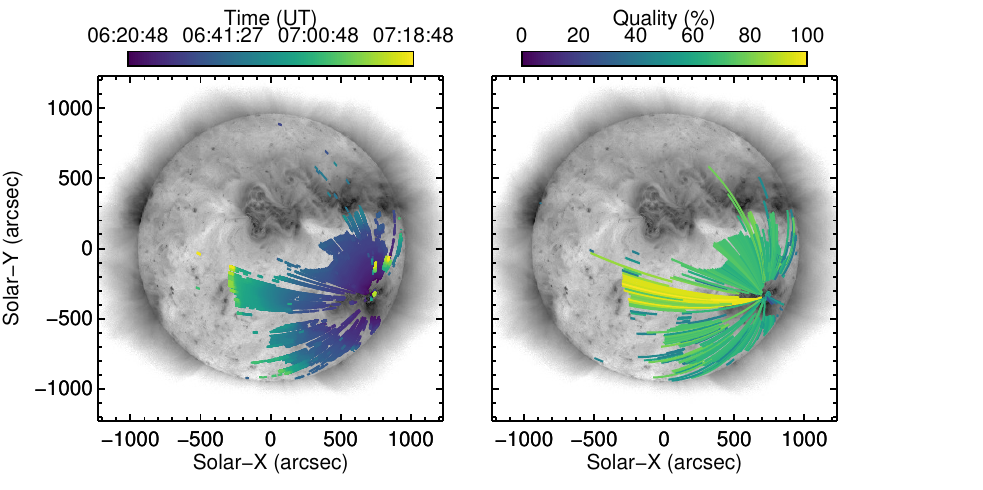}
    }
\caption{\emph{Panel~a}; Temporal evolution of the global wave associated with the solar eruption on 2011~June~7 
	      derived using the CorPITA code with colour showing time since 06:20:48~UT. \emph{Panel~b}; The quality rating
	      \citep[\cf][]{Long:2014} associated with each arc sector.}
\label{fig:corpita}
\end{figure*}

The events listed in the wave list of \citet{Nitta:2013} were processed using the Coronal Pulse Identification and Tracking 
Algorithm \citep[CorPITA;][]{Long:2014}. CorPITA is an automated code designed to identify, track and analyse global EUV 
waves using science quality data from \corrtwo{the} SDO/AIA 211~\AA\ passband. Although global waves have previously been 
characterised using each of the eight extreme ultraviolet (EUV) passbands observed by SDO \citep{Liu:2011}, previous 
work has shown that they are best observed using either the 193~\AA\ or 211~\AA\ passbands \citep{Long:2014}. The 
211~\AA\ passband was used here as it observes slightly hotter plasma than the 193~\AA\ passband and as a result does 
not have as much emission from the background corona, making the global wave easier to identify and characterise.

Percentage base difference images are used to identify the wave pulse, with the pre-event image defined as the 
image 2~minutes prior to the start time of the flare. A series of 360 arc sectors, each of 10$^\circ$ wide and offset by 
1$^\circ$, centered on the flare location are then used to define a series of intensity profiles. The wave pulse is identified 
using these intensity profiles \corr{and fitted using a Gaussian function which allows the position (\ie, the centroid of the 
Gaussian), peak intensity and full-width at half-maximum to be recorded in each arc sector for each time step. The CorPITA 
code then identifies the pulse \corrtwo{in each arc sector} by finding the largest section of contiguous data-points exhibiting 
increasing distance away from the source point. This section of data-points is then fitted using a quadratic function which 
provides an estimate of the initial pulse velocity and acceleration. The temporal variation in pulse distance from the source point} 
for the 2011~June~7 event is shown in Figure~\ref{fig:corpita}a. \corr{Although this technique ensures a consistent approach 
to estimating the initial pulse velocity and acceleration of the pulse, the accuracy can be affected by sudden jumps in pulse 
position (\eg, due to the algorithm becoming confused by bright points or small-scale loop oscillations). The accuracy of the 
measurement in each arc sector is therefore quantified in each case} using a quality rating system, with the pulse scored 
according to the number of images used to identify it, the fitted \corr{initial} velocity and acceleration and the uncertainty in 
identifying the pulse \citep[\cf][]{Long:2014}. Figure~\ref{fig:corpita}b shows the estimated quality rating for each of the arc 
sectors studied for the 2011~June~7 event. \corr{A wave is then identified by CorPITA if more than 10 adjacent arc sectors 
detect a moving pulse with a quality rating greater than 60\%.} \corrtwo{The measured parameters of the wave identified by 
CorPITA are then stored for future use (as listed in Table~\ref{table1}). In addition to the location of the source, start time of 
the wave and fitted kinematics, CorPITA also records the number of arcs in the largest segment in which CorPITA has identified 
a wave (Num.~arcs) and the central arc of this segment in degrees clockwise from solar north (Central arc angle). Note that 
Central arc angle uses the central angle of the segment to indicate the mean direction of the identified wave pulse, and does 
not necessarily correspond to the highest rated arc within that segment.} 

Although a more detailed description of the CorPITA technique may be found in the paper by \citet{Long:2014}, it should 
be noted that the code has since been updated to address issues found during an initial attempt at the work described here. 
Data is now downloaded in 20 minute chunks to speed up processing rather than an initial 10 minute chunk followed by 
single image downloading as before. This increase in time over which to search for a wave provides a better opportunity 
to identify the wave pulse, particularly for gradual flare events where the starting time of the wave and starting time 
of the flare may not be well correlated. The code has been rewritten for stability and to ensure a more rigorous analysis, 
while the colour table has also been updated to make it more accessible and easier to understand (as shown in
Figure~\ref{fig:corpita}). 

\subsection{Solar Energetic Particle Analysis}\label{ss:seps}

The SEP events associated with each global wave event were identified using measurements from the 3D EESA/PESA 
\citep{Lin:1995} instrument onboard the \emph{Wind} spacecraft. The data for electrons of energies between 
$\approx$1.3~keV and $\approx$27~keV and protons of energies between $\approx$195~keV and $\approx$4.4~MeV were examined for 24 
hours around (8 hours before and 16 hours after) the start time of the associated solar flare. In each case, the data 
were obtained from the NASA CDAW website\footnote{http://cdaweb.sci.gsfc.nasa.gov/index.html/}.

\begin{figure*}[!t]
\centering{
	\includegraphics[width=0.99\textwidth,clip=,trim=0mm 2mm 0mm 0mm]{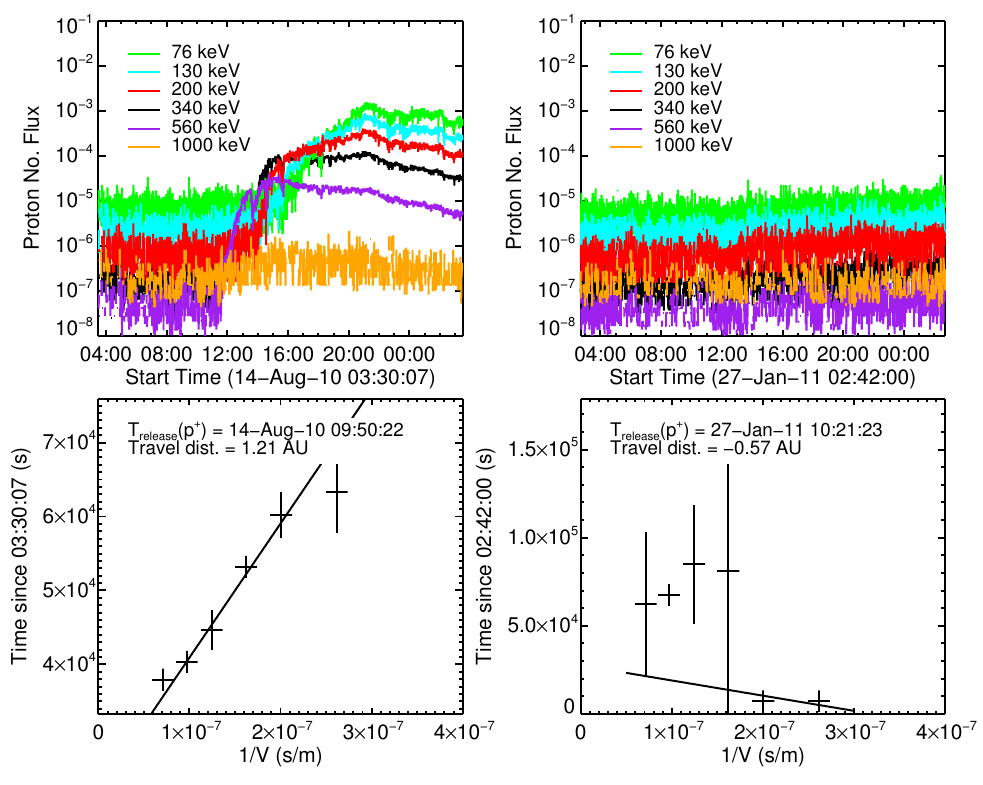}
    }
\caption{Proton flux (top row) and resulting velocity dispersion plots (bottom row) for a wave with an associated SEP 
		 event (2010~August~14; left column) and without an associated SEP event (2011~January~27; right column). SEP 
         onset is detected automatically in each case (see main text), with the onset time in each energy band used 
         to construct the velocity dispersion plot and estimate the time of particle acceleration on the Sun.}
\label{fig:sep_ex}
\end{figure*}

The presence of an SEP event in each energy band was determined by first smoothing the flux data using a Savitsky-Golay
filter to reduce the effect of small-scale variations. The flux was then examined to find the point at which it began to 
increase rapidly, with data prior to this point defined as the background. The onset times were then calculated 
using a Poisson cumulative sum (CUSUM) method \citep[\cf][]{HH:2005}. The CUSUM method is widely used in industry to 
identify changes in running processes, with a Poisson-CUSUM approach used if the data has a Poisson distribution. This 
approach works by cumulating the difference between an observed count $Y_i$ and a reference value,
\begin{equation}
k = \frac{\mu_d - \mu_a}{\textrm{ln}(\mu_d) - \textrm{ln}(\mu_a)},
\end{equation}
where $\mu_a$ is the mean of the background flux and $\mu_d$ is $\mu_a$ plus twice the standard deviation of the 
background flux. If this cumulation exceeds \corr{a} threshold value h \corr{(chosen to minimise the effect of small point-to-point 
fluctuations while retaining sensitivity to large-scale particle events)}, then an out of control signal is given. In this case, a particle 
event was identified by looking for 300 out of control signals in a row. The onset time of particle detection at the spacecraft was 
then defined by the time of the first out of control signal.

Once an onset time was determined for each energy level, a velocity dispersion analysis was used to determine the release 
time of the particles from the Sun \citep[\eg][]{Reames:2009}. This was also used to confirm the presence of an SEP event, 
with an event defined to have occurred if the velocity dispersion analysis gave a realistic physical release time from the 
Sun, \ie, the release time was before the onset time. An example of two events with realistic and unrealistic release times 
is shown in Figure~\ref{fig:sep_ex}. In the realistic case, the onset time at each energy can be identified by the sudden 
increase in particle flux, which corresponds to a valid estimate of release time and travel distance using the velocity 
dispersion plot. However, no such increase may be discerned in the unrealistic case, leading to the physically impossible 
estimates of release time and travel distance in the resulting velocity dispersion plot. A commonly used technique, it 
should be noted that velocity dispersion analysis assumes both that particles at all energies are released at the same time 
and that particle scattering is energy independent. However, energy dependent particle scattering can greatly affect 
particle arrival times, meaning that propagation through the heliosphere is not a simple trajectory along the Parker Spiral.
Therefore, to ensure consistency, each event was initially processed using this approach with the resulting plots examined 
`by-eye' for confirmation.

\subsection{Identification and Characterisation of Type~\Rmnum{2} Radio Bursts}\label{ss:radio}

The type~\Rmnum{2} radio bursts associated with each global wave event were identified using the daily lists of radio 
bursts collated by the Space Weather Prediction Centre located at the National Oceanic and Atmospheric Administration 
(NOAA/SWPC). A window 90~minutes either side of the start time of the global wave was used to look for associated events 
in the NOAA/SWPC list. This choice of time window was motivated by the fact that type~\Rmnum{2} radio bursts can be seen 
up to 15~minutes before or after the first instance of an EUV wave observation \citep{Park:2013,Warmuth:2010,Miteva:2014}; 
the larger time window used here was chosen to account for any anomalous events. For each candidate radio burst associated 
with a global wave, the SWPC list provides an associated start time, end time, observatory used to make the observation, 
frequency range of the burst and estimated drift speed. It should be noted that the drift speeds used in this analysis 
are the values provided by NOAA/SWPC, which are obtained using the standard approach of converting drift rate to speed 
via a density model of the solar corona \citep{Mann:1995,Mann:2007}. However, it should be remembered that the absolute
values of these bursts are known to be subject to large uncertainty given the often arbitrary nature of the models \corr{and the differences between individual observatories} 
\citep{Magdalenic:2008, Magdalenic:2012}. This estimation also assumes a radially propagating shock driver, whereas it 
has been shown that shocks can often propagate non-radially \citep{Mancuso:2004,Magdalenic:2012}. Nonetheless, given 
the goal of trying to find a statistical correlation between radio burst properties and various other phenomena, we 
believe that the uncertainties associated with using the quoted type~\Rmnum{2} burst speeds should be acknowledged but 
are not of major concern.

\section{Results}\label{s:results}      

Due to limitations inherent to the approach taken by CorPITA, it was not possible to analyse all of the events identified 
by \citet{Nitta:2013} as they originated either too close to or beyond the solar limb. CorPITA requires a source point 
from which to track the pulse and so it cannot study events that do not originate on disk. As a result, of the 410 wave 
events identified by \citet{Nitta:2013}, only 362 could be analysed using CorPITA. 164 events were classified as having 
global waves by CorPITA, with no waves found for the remaining 198 events. The output from CorPITA for all events studied 
is listed in Table~\ref{table1}.

\subsection{Wave kinematics}\label{ss:kins}

\begin{figure*}[!t]
\centering{
	\includegraphics[width=0.99\textwidth,clip=,trim=0mm 0mm 0mm 0mm]{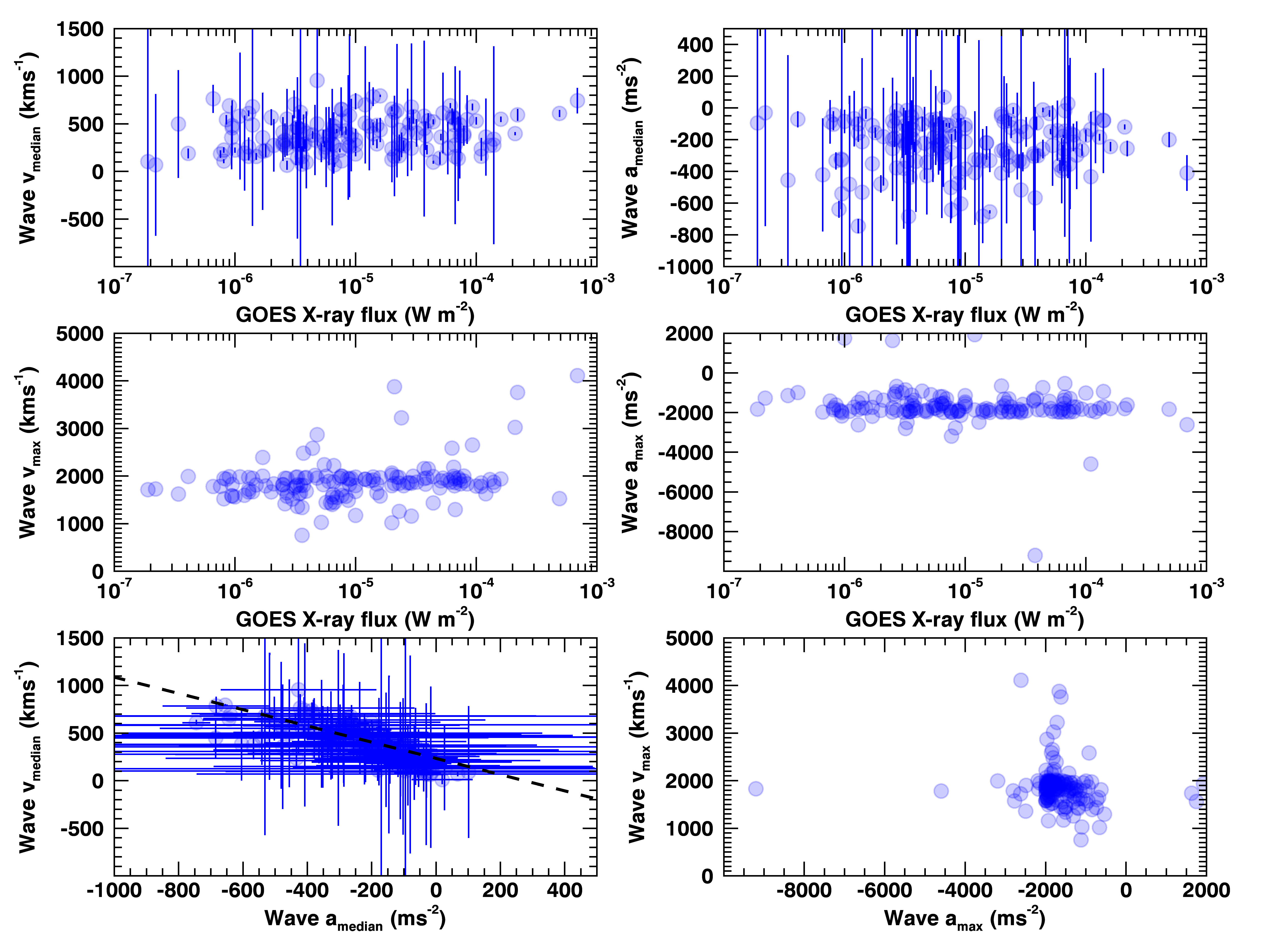}

    }
\caption{Relationship between the GOES X-ray flux of the associated flare and the median initial velocity (\emph{panel~a}), 
	     median acceleration (\emph{panel~b}), maximum initial velocity (\emph{panel~c}) and absolute maximum acceleration
	     (\emph{panel~d}) of the wave measured by CorPITA. Bottom two panels show the relationship between the median 
	     velocity and acceleration of the wave (\emph{panel~e}) and maximum velocity and acceleration of the wave 
         (\emph{panel~f}).}
\label{fig:wave_kins}
\end{figure*}

The global waves identified here exhibited a wide variety in their kinematics, both from arc-to-arc within each event and 
from event-to-event. For a given event, CorPITA is designed to examine each arc sector separately, allowing the directional 
variation in pulse position to be identified and studied. Although this provides a more accurate estimate of how the wave 
evolves, it makes event-to-event comparison difficult as it raises the question of which velocity and acceleration values 
should be used. \corrtwo{For every event studied here, the initial velocity and acceleration were calculated for each arc by 
first identifying the largest section of contiguous data-points exhibiting increasing distance away from the source point and 
fitting these points using a quadratic function \citep[as illustrated in Figure~3 of][]{Long:2014}. The median of the initial velocity 
and acceleration} values across all arc sectors with a sufficiently high quality rating \corrtwo{were then} chosen as being most 
representative of the kinematics of that event. Panels a \corr{and} b of Figure~\ref{fig:wave_kins} show the median \corr{initial} 
velocity and acceleration respectively of all events studied plotted with respect to the peak GOES X-ray flux of the associated 
flare in each case. It is clear that there is a broad spread in both the median \corr{initial} velocity and acceleration of the waves 
studied, with the maximum median \corr{initial} velocity peaking at $\approx$950~km~s$^{-1}$ and the maximum median 
acceleration peaking at $\approx -750$~m~s$^{-2}$.

To determine whether the median \corr{initial} velocity and acceleration of the waves was most appropriate for comparing 
events, the maximum \corr{initial} velocity and acceleration of the wave events were also examined. \corrtwo{These were 
taken as the maximum of the initial velocity and acceleration values derived across all arc sectors with a sufficiently high 
quality rating for a given event.} Panels c \corr{and} d of Figure~\ref{fig:wave_kins} show the maximum \corr{initial} velocity 
and acceleration respectively of all events studied plotted with respect to the peak GOES X-ray flux of the associated flare in 
each case, similar to panels a \corr{and} b of Figure~\ref{fig:wave_kins}. Although the maximum \corr{initial} velocity and 
acceleration are clustered around $\approx$2000~km~s$^{-1}$ and $\approx -$2000~m~s$^{-2}$ respectively, there is a 
much larger spread in both as apparent from panels c \corr{and} d of Figure~\ref{fig:wave_kins}. 

Finally, the \corr{initial} velocity and acceleration were plotted against each other for both the median and maximum values 
to try and identify any trends comparable to those previously found by \corr{\citet{Warmuth:2010}}, \citet{Warmuth:2011} 
\corr{and \citet{Muhr:2014}}. Panel~e of Figure~\ref{fig:wave_kins} 
shows the median \corr{initial} velocity plotted against median acceleration for each event studied. It is possible to identify a 
clear trend in this case, with faster (slower) waves exhibiting a stronger (weaker) negative acceleration. \corr{This is consistent 
with the results of both \citet{Warmuth:2011} and \citet{Muhr:2014} who plotted the initial velocity against acceleration as well as 
with \citet{Warmuth:2010} who plotted the average velocity against acceleration. The approach} was repeated 
for the maximum \corr{initial} velocity and acceleration (shown in Figure~\ref{fig:wave_kins}f), but consistent with panels~c 
\corr{and} d, the plot shows a much broader spread. Although a slightly decreasing trend can be discerned, with faster (slower) 
waves again exhibiting a stronger (weaker) negative acceleration, this trend is much weaker than that apparent for the median 
\corr{initial} velocity and acceleration data shown in panel~e. This suggests that the maximum \corr{initial} velocity and acceleration 
are poor indicators of the overall wave kinematics and the median \corr{initial} velocity and acceleration should be used when 
trying to characterise a given wave event using a single kinematic value.

Previous work by \citet{Warmuth:2011} used the wave catalogue of \citet{Thompson:2009} to suggest that there were three 
kinematic classes of global EUV waves. Class~1 referred to initially fast waves with a strong deceleration, class~2 waves 
had moderate and nearly constant speeds while class~3 referred to slow waves with an erratic kinematic profile. The results 
shown in Figure~\ref{fig:wave_kins}e are generally consistent with the results of \citet{Warmuth:2011}, but it is not possible to 
distinguish three independent kinematic classes of global waves in this case. This may be due to the larger number of events 
used \citep[164 here compared to 61 for][]{Warmuth:2011} or alternatively the higher cadence of SDO/AIA compared to 
SOHO/EIT and STEREO/EUVI in \citet{Warmuth:2011}. However, when the data in Figure~\ref{fig:wave_kins}e were fitted 
using a linear relation, an intercept of 180~km~s$^{-1}$ was found, consistent with the cutoff value of $\approx$170~km~s$^{-1}$ 
used by \citet{Warmuth:2011} to define the difference between linear waves and those features possibly due to magnetic 
reconfiguration. However, this value was slightly lower than the intercept found by \citet{Muhr:2014} using STEREO/EUVI 
observations. \corr{The consistency between the results presented here and by \citet{Warmuth:2010}, \citet{Warmuth:2011} and 
\citet{Muhr:2014} indicates that these features are large amplitude events as outlined by \citet{Long:2017}.}

\corr{Although the catalogue of \citet{Nitta:2013} was used as a starting point for this work, it is worth noting that the analysis of the 
pulse kinematics described here and shown in Figure~\ref{fig:wave_kins} do not match the results presented in \citet{Nitta:2013}. 
This is most likely due to the different approaches used to identify the wave pulse and measure its kinematics. Whereas \citet{Nitta:2013} 
tracked the leading edge of the wavefront in a 2-dimensional time-distance plot, here a 1-dimensional intensity profile of the wavefront 
was fitted using a Gaussian, with the centroid of the Gaussian taken as the position of the wavefront at each point in time. This 
approach removes the effect of pulse broadening on the identified pulse position \citep[see, \eg][for a more detailed discussion]{Long:2014}. 
In addition, both the velocity and acceleration of the pulse were simultaneoulsy measured here for each pulse using a single quadratic 
fit to the temporal variation in pulse position. This is in contrast to \citet{Nitta:2013}, who used both linear and quadratic fits independently 
applied to the temporal variation in pulse position to estimate the velocity and deceleration of the pulse respectively.}

\subsection{Relationship with solar flares and active regions}\label{ss:flare}

\begin{figure*}[!t]
\centering{
	\includegraphics[width=0.99\textwidth,clip=,trim=0mm 5mm 0mm 0mm]{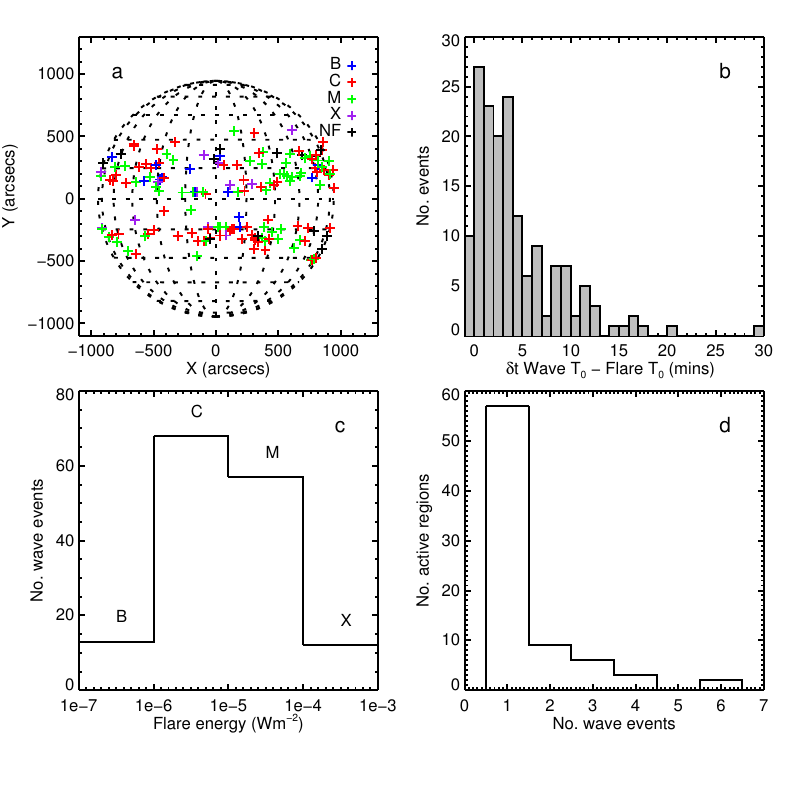}
    }
\caption{\emph{Panel~a}; The location of every flare with an associated wave identified by CorPITA with colour indicating 
	      flare class. \emph{Panel~b}; The relationship between the start time of the flare as defined by GOES and the 
	      \corr{time of the first wavefront observed} by CorPITA. \emph{Panel~c}; The number of global waves associated with 
	      each flare class. \emph{Panel~d}; The relationship between active regions and global waves.}
\label{fig:flare_relation}
\end{figure*}

When global ``EIT waves'' were first observed there was a lot of discussion regarding their origin, with the debate focusing 
on whether they were initially driven by the associated flare or coronal mass ejection \citep[\eg,][]{Cliver:2005,Vrsnak:2006}. 
Since then, a general consensus has been reached that they are initially driven by the rapid lateral expansion of the erupting 
CME in the low corona; a conclusion strongly supported by the SDO observations reported by \citet{Patsourakos:2010}. A 
detailed discussion of the predictions made by the different theories and how recent observations support this conclusion may 
be found in the recent paper by \citet{Long:2017}. 

As shown in panel~a of Figure~\ref{fig:flare_relation}, all of the events studied here originated in the activity belts, consistent 
with previous observations \citep[\eg,][]{Muhr:2014}. However, 14 of the events studied had no associated flare. The vast 
majority of the waves also tended to start after the start time of the associated flare, as indicated by panel~b of 
Figure~\ref{fig:flare_relation}, although 10 events studied did start before the flare. While most of the waves identified \corr{were 
first observed by CorPITA} within 10 minutes of the start of the flare, some \corr{were not observed} until up to 30~minutes after 
the flare (as defined by the NOAA GOES catalogue). Panel~c of Figure~\ref{fig:flare_relation} indicates a broad spread in the 
size of the flares associated with each wave studied. Although the vast majority of waves were associated with M- and C-class 
flares, more than 10 waves were associated with both B- and X-class flares respectively. 

These results suggest that the flare plays little-to-no role in the initiation or even existence of a wave, indicating that some other 
criteria must be fulfilled before a solar eruption produces a wave. While the relationship with CMEs is examined more closely in 
Section~\ref{ss:cme}, panel~d of Figure~\ref{fig:flare_relation} suggests that the active region from which the wave originates 
may be important. Over 25~\% of active regions produced more than one wave during their time on-disk, with two active regions 
producing 6 waves each. This suggests that the magnetic structure of the active region or its relationship with the surrounding 
quiet solar corona may determine the ability of the active region to produce a wave during a solar eruption.

\subsection{Relationship with CMEs}\label{ss:cme}

\begin{figure*}[!t]
\centering{
	\includegraphics[width=0.99\textwidth,clip=,trim=0mm 0mm 0mm 0mm]{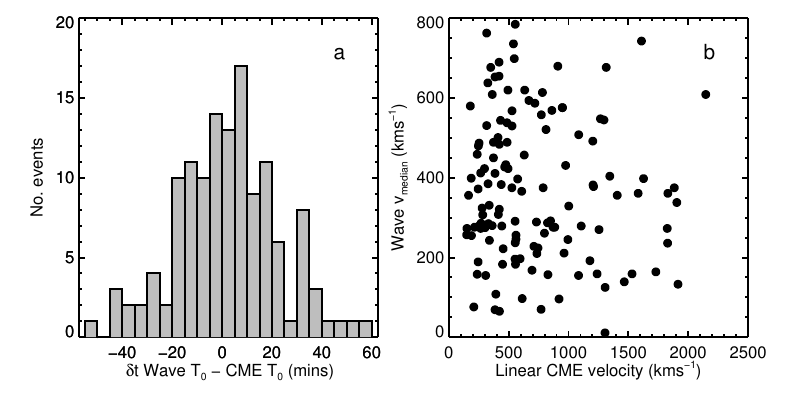}
    }
\caption{\emph{Panel~a}; The relationship between the start time of the CME as defined by the \corr{linear fit to the temporal 
              variation in CME position obtained from the} LASCO CDAW catalogue and the start time of the wave detected by 
              CorPITA. \emph{Panel~b}; The relationship between the median initial wave velocity measured by CorPITA and the 
              linear starting velocity of the CME measured by CDAW.}
\label{fig:cme_relation}
\end{figure*}

A comparison was also made between the identified waves and their associated coronal mass ejections as identified by the LASCO 
CDAW catalogue\footnote{\url{https://cdaw.gsfc.nasa.gov/CME_list/}}. Global waves have historically been strongly associated 
with CMEs, with \citet{Biesecker:2002} in particular suggesting that every wave has an associated CME while not every CME has 
an associated wave. However, a direct comparison between the on-disk global wave and the erupting CME is complicated by the 
fact that CMEs are best seen when in the plane of the sky (so when associated with eruptions close to the limb) whereas global 
waves are best observed and tracked on-disk.

As shown in Figure~\ref{fig:cme_relation}, there is no clear correlation between the start time of the global wave and the start time 
of the associated CME predicted by the \corr{linear fit to the temporal variation in CME position obtained from the} LASCO CDAW 
catalogue (panel~a). Similarly, there is no clear correlation 
between the median initial wave velocity and the fitted linear velocity of the CME (panel~b). While initially concerning, both 
of these results are consistent with our current understanding of global waves and CMEs and their relationship; a point worth 
discussing in more detail.

The CDAW catalogue uses LASCO observations to identify and measure CMEs and, as a consequence of the location of SOHO/LASCO at 
the L1 Lagrange point, is biased towards CMEs erupting from the solar limb as seen from Earth. As a result, CMEs associated 
with global waves observed by SDO/AIA tend to be observed as halo CMEs, which are difficult to identify and measure. The velocity 
and starting time values used here were also taken from the linear fits to the temporal variation of the CME distance, with the
result that any acceleration or deceleration of the CME was ignored, which may account for the spread in projected CME start 
times. Finally, the predicted mechanism by which the waves are produced by the lateral expansion of the erupting CME in the low
corona suggests that there should be no direct correlation between the velocity of the wave and the forward velocity of the CME
(which is what is typically measured when studying CMEs). As a result the lack of any correlation shown in
Figure~\ref{fig:cme_relation} is to be expected.

\subsection{Relationship with Type~\Rmnum{2} radio bursts}\label{ss:rad_rel}

\begin{figure*}[!t]
\centering{
	\includegraphics[width=0.98\textwidth,clip=,trim=0mm 2mm 0mm 0mm]{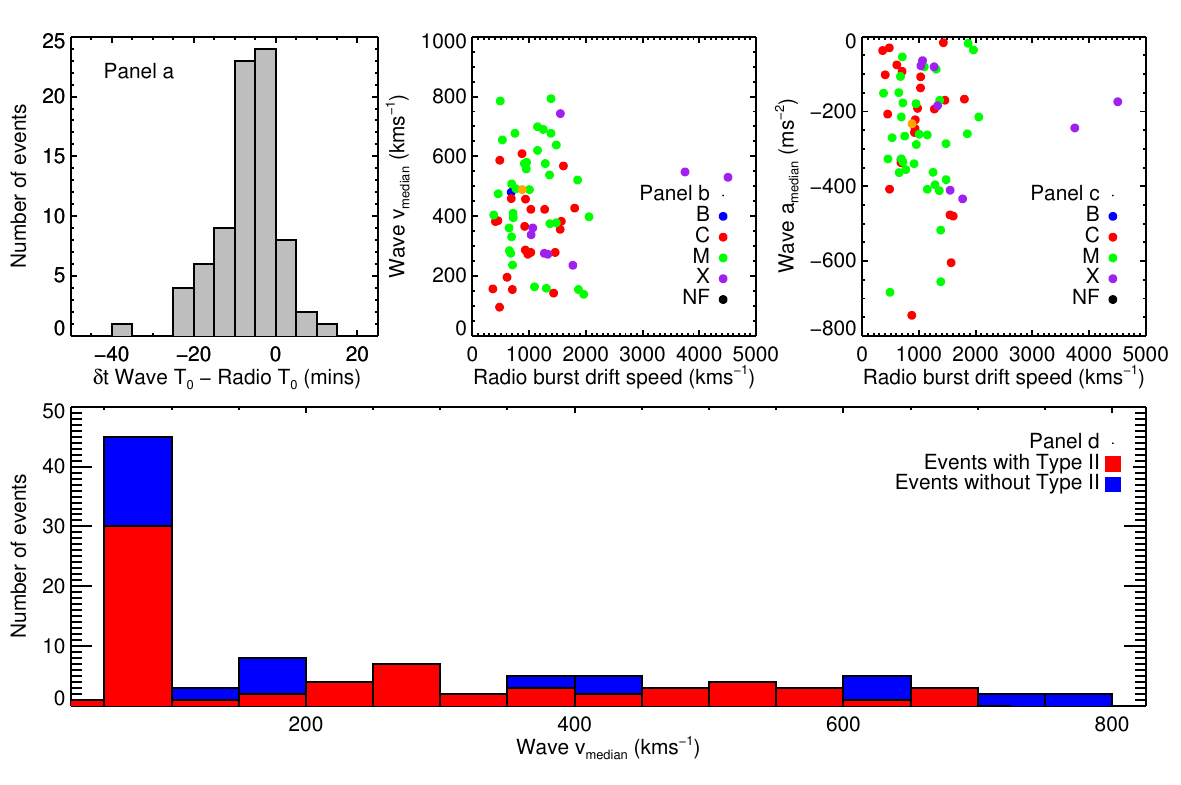}
    }
\caption{\emph{Panel~a}; The offset time between the start of the global wave and the associated type~\Rmnum{2}
	      radio burst. The relationship between the median velocity (\emph{Panel~b}) and median acceleration 
          (\emph{Panel~c}) of the wave as measured by CorPITA and the drift speed of the type~\Rmnum{2} radio 
          burst. \emph{Panel~d}; Histogram showing the variation in median wave velocity for wave events with 
          and without type~\Rmnum{2} radio bursts.}
\label{fig:radio_relation}
\end{figure*}

With the growing consensus on the interpretation of global coronal waves as large-amplitude waves initially driven by the 
lateral expansion of a CME in the low corona, a natural comparison can be made between global waves and type~\Rmnum{2}
radio bursts. Type~\Rmnum{2} bursts are strongly associated with MHD shock waves \citep[\cf][]{Nelson:1985} and as a result 
the relationship between them and global waves has long been hypothesised and investigated 
\citep[\eg,][]{Cliver:1999,Biesecker:2002,Nitta:2014}. However, their relationship remains inconclusive. 

Despite the strong relationship between type~\Rmnum{2} radio bursts and MHD shock waves and the interpretation of global EUV 
waves as large-amplitude or shock waves, a comparable number of wave events were identified with and without associated radio 
bursts. 66 wave events had an associated type~\Rmnum{2} burst, with 98 wave events having no associated type~\Rmnum{2} radio 
signature detected at Earth, meaning that 40~\% of the waves in our sample have an associated type~\Rmnum{2} burst. This is 
higher than the 22~\% association rate reported by \citet{Muhr:2014}, but comparable to the 43~\% association rate reported by 
\citet{Biesecker:2002} for waves with a high quality rating \citep[$>$50~\% using the classification of][]{Thompson:2009}. 
\corr{However, this is is much lower than the 100~\% associated rate between type~\Rmnum{2} bursts and EUV waves with an 
associated H-$\alpha$ Moreton-Ramsey wave reported by \citet{Warmuth:2004} and \citet{Warmuth:2010}. It should be noted that 
no comparison was made in this paper between global waves observed in EUV and H-$\alpha$ passbands; given the relative lack of 
recent synoptic studies of Moreton-Ramsey waves in H-$\alpha$ data we leave a more detailed analysis of the relationship between 
these two phenomena for a dedicated future work.}

As shown in Figure~\ref{fig:radio_relation}d, there is no relationship between the median velocity of the wave and whether 
or not it had an associated type~\Rmnum{2} burst, with very fast wave events exhibiting no type~\Rmnum{2} emission while a
significant number of very slow events had associated radio emission. This is most likely related to the fact that 
type~\Rmnum{2} radio burst generation is related to conditions in the upper corona at $>1.2$~R$_{\odot}$ \citep{Mann:2003}, 
while the global wave propagates lower down in the corona \citep[at $\approx 70$--100~Mm, 
\cf][]{Kienreich:2009,Patsourakos:2009a}.

For those events which had an associated type~\Rmnum{2} radio burst, the start of the radio emission was observed after the 
start of the wave in the vast majority of cases (see Figure~\ref{fig:radio_relation}a), similar to the relative start times 
of waves and radio bursts reported by \citet{Miteva:2014} and \citet{Warmuth:2010}. This delay between the start time of the 
global wave and the associated type~\Rmnum{2} burst is likely due to the time taken for the disturbance to either become 
super--Alfv\'{e}nic \citep[as modeled by][]{Vrsnak:2000}, or the time taken for the driver to reach regions of low ambient 
Alfv\'{e}n speed in the corona \citep{Mann:2003,Zucca:2014}; in some cases this can take up to 30~minutes after detection of 
the wave. \corr{The duration of the radio bursts was also observed to be at most 35~minutes, with most radio bursts lasting less 
than 20~minutes, consistent with the lifetime of the observed waves.} 
However, a number of events had associated radio emission which was observed to start prior to the first detection 
of the global wave, suggesting that the radio emission in those cases may have been due to either the rapid expansion of the 
CME rather than the wave, or the CME driving a shock radially before any lateral expansion produced the wave. In either case, 
the difference in start time of up to 30~minutes between the waves and associated type~\Rmnum{2} radio bursts means that while 
they may originate from a common MHD disturbance in the corona, they most likely belong to spatially separated parts of 
this disturbance.

There is also no clear relationship between either the median initial velocity or acceleration of the identified waves and the 
drift speed of the associated radio bursts. Panels~b \corr{and} c of Figure~\ref{fig:radio_relation} show that most of the events 
identified tended to have drift velocities of 0-2000~km~s$^{-1}$, with two notable exceptions in both cases. Some of the events 
also had quite high drift velocities of $\approx$2000~km~s$^{-1}$ despite estimated median velocities lower than 200~km~s$^{-1}$.
Even when accounting for the size of the associated flare, there is still no clear relationship between median initial wave 
velocity and drift speed of the associated type~\Rmnum{2} burst, with the flare size in panels b \corr{and} c of
Figure~\ref{fig:radio_relation} indicated by the colour of the data-point.

This lack of a clear relationship between the drift velocity of the type~\Rmnum{2} burst and the median initial velocity of 
the identified wave is at odds with the conclusions of \citet{Warmuth:2010}, who found a linear relationship between these
parameters. \corr{Several individual case studies have also reported a kinematical relationship between type IIs 
\citep[\eg,][]{Vrsnak:2006,Pohjolainen:2008,Grechnev:2011,Kozarev:2011,Ma:2011}. However, there may be several reasons 
for the lack of such a relationship that we find here. Firstly}, as mentioned above, while the type~\Rmnum{2} radio 
burst and the global wave are most likely different manifestations of the same shock in the corona, they may belong to separate parts 
of this shock, \ie, the wave propagates laterally or parallel to the solar surface, while the type~\Rmnum{2} burst may propagate
(semi-)radially \citep{Grechnev:2011}. Second, there is no guarantee that these speeds will be the same. In fact, the discrepancy
between the type~\Rmnum{2} drift speed and the wave speed means that expansion of the MHD disturbance is most likely anisotropic,
with no relationship between lateral and radial expansion speed. \corr{Thirdly}, a major issue with relating these two speeds concerns 
the reliability of the type~\Rmnum{2} speed itself. This speed is derived from one of the many density models used in radio physics, 
which are often chosen arbitrarily and may not represent the density environment of the event \citep{Magdalenic:2008}. This analysis 
also assumes radial radio source propagation, which may not be the case. \corr{Finally, \citet{Warmuth:2010} focussed on global EUV 
waves with associated H-$\alpha$ Moreton-Ramsey waves. Given the supposed mechanism by which these phenomena are thought 
to be related \citep[\ie, the coronal EUV wave has a sufficiently large downward impulse that allows its footprint to be observed as a 
Moreton-Ramsey wave, see, \eg][for more details]{Warmuth:2010,Warmuth:2015}, this indicates that the waves observed by 
\citet{Warmuth:2010} were large-amplitude shocks. Although a similar mechanism may have produced the global waves studied here, 
these waves may not have been sufficiently strong to produce a radio signature. This suggests that a more granular study, focussing 
on strong EUV waves with associated H-$\alpha$ Moreton-Ramsey waves might provide a higher correlation between global EUV waves 
and type~\Rmnum{2} radio emission \citep[comparable to][]{Warmuth:2004,Warmuth:2010}.}

\subsection{Relationship with Solar Energetic Particle events}\label{ss:sep_rel}

Similar to the predicted relationship with type~\Rmnum{2} radio bursts, as large-amplitude and sometimes weakly shocked 
waves, global waves would be expected to accelerate solar energetic particles as they propagate across the Sun. However, 
of the 164 events with identified global waves, only 21 were found to have any evidence of an associated proton event, 
whereas only 14 events were found to have any evidence of an associated electron event (of which 12 had an associated 
proton event). While there are several probable reasons for this discrepancy, it is most likely due to a lack of 
connectivity between the field lines along which the particles could be accelerated and the spacecraft detecting the 
particles. In a simplistic interpretation, only events erupting from solar west would be expected to have any connectivity 
with the detecting spacecraft due to the Parker Spiral, a suggestion consistent with the fact that 19 of the 23 events 
found here were associated with flares that originated on the western hemisphere.

However, recent studies have shown that even events on the far side of the Sun can produce particle detections at Earth, 
with studies by \citet{Rouillard:2012} and \citet{Lario:2016} suggesting that the global waves traveled across the Sun 
and eventually encountered field lines connected to Earth/L1. It has also been shown that particles can be detected despite 
being produced far from the footpoint of a magnetic field line connected to the spacecraft as a result of either super-radial
spreading of field lines throughout the corona \citep[\eg][]{Klein:2008}, or very high levels of lateral diffusion in a 
turbulent solar wind \citep[\eg][]{Laitinen:2016}. The large spatial extent of global waves and their ability to accelerate 
particles far from the erupting active region therefore suggests that more SEP events should have been identified. However, 
only 4 events \corr{were} found here that were associated with active regions located in the eastern solar hemisphere. 

This lack of identified events may be due to the configuration of the coronal magnetic field into which the wave is 
propagating. Although previous work by \citet{Park:2013,Park:2015} compared wave propagation to a Potential Field Source 
Surface extrapolation of the solar corona to compare the time at which the wave encountered a connected field line with the 
inferred SEP release times, this approach is rarely taken. However, the results described here suggest that a full understanding 
of the ability of a global wave to accelerate SEPs must combine the propagation of the wave with a full understanding of the 
coronal and heliospheric magnetic field. Although this approach is beyond the scope of the statistical path taken here, we hope 
to return to it in a dedicated future work.

\section{Discussion and Conclusions}\label{s:disc}      

In this manuscript, we have applied the Coronal Pulse Identification and Tracking Algorithm to the list of global coronal 
waves assembled by \citet{Nitta:2013} and compared the output with a variety of other solar phenomena. Of the 410 events
identified as waves by \citet{Nitta:2013}, only 362 could be processed using this approach due to requirements on the location
of the source point of the wave. Of these 362 events processed, 164 were classified as having associated global waves, with 
CorPITA finding no waves for the remaining 198 events. This indicates a significant disconnect between the systematic automated
approach to identifying and characterising global coronal waves and the traditional `by-eye' approach. Although most of these 
issues can and should be overcome through advanced image and signal processing and feature tracking, some may be due to the 
different definitions used to identify the wave pulse. For example, CorPITA uses a series of 1-dimensional intensity profiles
obtained from percentage-base difference images to identify and fit the wave using a Gaussian model, whereas \citet{Nitta:2013} 
used 2-dimensional distance-time stack plots to identify the leading edge of the pulse. 

With the waves identified and tracked, the next step was to examine the variation in wave kinematics. As CorPITA uses a series 
of 360 10$^\circ$ arc sectors to identify and track the waves, each wave can be represented by a range of velocity and 
acceleration values which may not be representative of the large-scale motion of the wave. The maximum velocity and acceleration 
in particular are poorly representative of the overall wave motion as they can be strongly affected by anomalous measurements.
However, the median velocity and acceleration provide a much better representation of the large-scale kinematics and should 
always be used when wishing to describe a wave using a single kinematic value. The median velocity and acceleration of the 
measured waves were also found to be correlated \citep[consistent with the previous work of][]{Warmuth:2011}. Although the 
spread in values increased with both velocity and acceleration, faster waves tend to have a stronger deceleration. However, 
no clear relationship could be determined for the maximum velocity and acceleration, suggesting that they are not representative 
of the overall kinematics of the wave.

There was also no clear relationship between the global waves and the associated solar flares. Neither the maximum or median
velocities or accelerations showed any relationship with the class of the associated flare, with C-class flares associated 
with the waves that exhibited both the highest and lowest median velocity. Similarly, there was no clear correlation between 
the wave parameters described here and the properties of the associated CMEs as measured by the LASCO CDAW catalogue. However 
this is most likely due to instrumental and measurement effects rather than the lack of any underlying physical relationship 
and isunsurprising given the statistical approach taken here. A more thorough analysis would require measurements of CMEs 
taken away from the Sun-Earth line and would measure the lateral expansion velocity rather than the forward motion of the CME. 
It was possible to observe CMEs propagating along the Sun-Earth line using the STEREO spacecraft, with the instruments on both
spacecraft deliberately designed to have  overlapping fields-of-view in the low corona which could allow the lateral expansion 
of the CME to be measured. However, the low temporal cadence of the instruments makes it difficult to disentangle the wave 
from the expanding CME, while the gradual progression of STEREO behind the Sun during the period of this project complicates 
a direct comparison between the global wave and the associated CME. Finally, no correlation was found between the global waves 
and SEPs, with only 21 of the events exhibiting any SEP signature. This is most likely because only \emph{Wind} data were used 
in this analysis, but the ability of global waves to accelerate particles far from their erupting active region  suggests that 
the structure of the coronal magnetic field into which the wave propagates should be accounted for when trying to study the
acceleration of SEPs by global waves in the solar corona.

The lack of any clear statistical correlation between the different solar phenomena studied here indicates that determining the criteria 
required to produce a global wave is not a simple task. Although the majority of waves identified here were associated with 
both flares and CMEs, the parameters measured show no correlation, suggesting that the free energy available within an active 
region to produce a flare and accelerate the CME does not determine the presence of a wave in the subsequent eruption. However, 
it was found that over 25~\% of active regions produced multiple wave events, with 2 active regions producing 6 wave events 
each. This suggests that the structure of the erupting active region and its connectivity with the surrounding quiet solar 
corona may be much more important for determining the presence of a global wave in a solar eruption. Understanding the criteria 
required to produce a global wave therefore requires a more detailed examination of an active region producing multiple wave 
events, which we hope to continue in a dedicated future work.

\begin{acks}
The authors wish to thank Alexander~Warmuth for useful discussions \corr{and the anonymous referee 
whose comments helped to improve the paper}. DML is an Early Career Fellow funded by the Leverhulme 
Trust. PM acknowledges a grant awarded by the SCOSTEP/VarSITI consortium. GG is supported by a UCL 
IMPACT studentship. EPC is supported by ELEVATE: Irish Research Council International Career 
Development Fellowship - co-funded by Marie Curie Actions. DPS was funded by the US Air Force Office 
of Scientific Research under Grant No.~FA9550-14-1-0213. \corrtwo{Data from SDO/AIA is courtesy of 
NASA/SDO and the AIA science team.}
\end{acks}

\begin{footnotesize}
\noindent
\textbf{Disclosure of Potential Conflicts of Interest}  The authors declare that they have no conflicts of interest.
\end{footnotesize}

\appendix

\corr{Table~\ref{table1} below outlines the complete list of global wave events identified by \citet{Nitta:2013} and 
processed by the CorPITA code for this paper. \emph{t\_flare} refers to the start time of the flare as defined by the 
GOES classification. \emph{t\_wave} refers to the time of the first observation of the wave by CorPITA.} 
\emph{Num.~arcs} refers to the number of arcs in the largest segment in which CorPITA has identified a wave, with 
\emph{Central~arc} referring to the central arc of this segment \corr{in degrees clockwise from solar north}. 
Median velocity is given in km~s$^{-1}$ and median acceleration is given in m~s$^{-2}$. \corr{The complete list is also} 
available as a Comma Separated Value list attached to the online version of this paper. 

\begin{landscape}
\begin{centering}
\footnotesize

\pgfplotstabletypeset[
      begin table=\begin{longtable},
      begin table/.add={}{[!t]},
      header=true,
      trim cells,
      column type = c,
      multicolumn names, 
      columns/Flare_date/.style={column name=Date, {string type}},
      columns/x/.style={column name= Flare X, precision=2}, 
      columns/y/.style={column name= Flare Y, precision=2},
      columns/Flare_st/.style={column name=t$_{flare}$, {string type}},
      columns/wave_time/.style={column name=t$_{wave}$, {string type}},
      columns/Flare_class/.style={column name=Flare, {string type}},
      columns/Active_Region/.style={column name=Active, {string type}},
      columns/Num_consecutive_arcs/.style={column name=Num., {string type}},
      columns/Central_arc_angle/.style={column name=Central, {string type}},
      columns/Wave_present/.style={column name=Wave, {string type}},
      columns/median_vel/.style={column name=Median, {string type}},
      columns/median_acc/.style={column name=Median, {string type}},
      every head row/.append style={before row={\caption{CorPITA analysis.}\label{table1} \\ \midrule}, 
       after row={ & arcsec & arcsec & UT & UT & class & Region & arcs & arc &  & initial vel. & acc. \\
       \midrule\endfirsthead}},
      every first row/.style={before row={\multicolumn{12}{c}{\textbf{\tablename\ \thetable{}} -- 
       \textit{continued from previous page}} \\ \midrule
      Date & Flare X & Flare Y & t$_{flare}$ & t$_{wave}$ & Flare & Active & Num. & Central & Wave & Median & Median \\
        & arcsec & arcsec & UT & UT & class & Region & arcs & arc & & initial vel. & acc. \\ \midrule\endhead}},
      every last row/.append style={after row=\bottomrule},
      end table=\end{longtable}
    ]{CorPITA_database.txt}
\end{centering}
\end{landscape}

\bibliographystyle{spr-mp-sola}
\bibliography{wave_sep_bib_file}  

\end{article} 

\end{document}